\documentclass[aps,prl,preprint,groupedaddress]{revtex4-1}

\usepackage{graphicx}
\begin{document}


\title{Anomalous Quantum Oscillations of Interacting Electron-hole Gases in Inverted Type-II InAs/GaSb Quantum Wells}


\author{Di Xiao}
\affiliation{Department of Physics, the Pennsylvania State University, University Park, PA 16802, USA}
\author{Lun-Hui Hu}
\affiliation{Department of Physics, Zhejiang University, Hangzhou, Zhejiang, 310027, China}
\author{Chao-Xing Liu}
\author{Nitin Samarth}
\email[]{nsamarth@psu.edu}
\affiliation{Department of Physics, the Pennsylvania State University, University Park, PA 16802, USA}



\date{\today}

\begin{abstract}
We report magneto-transport studies of InAs/GaSb bilayer quantum wells in a regime where the interlayer tunneling between the electron and hole gases is suppressed. When the chemical potential is tuned close to the charge neutrality point, we observe anomalous quantum oscillations that are inversely periodic in magnetic field and that have an extremely high frequency despite the highly insulating regime where they are observed. The seemingly contradictory coexistence of a high sheet resistance and high frequency quantum oscillations in the charge neutrality regime cannot be understood within the single-particle picture. We propose an interpretation that attributes our experimental observation to the Coulomb drag between the electron and hole gases, thus providing strong evidence of the significance of Coulomb interaction in this topological insulator. 
 
\end{abstract}


\maketitle


The interplay between topology and interactions in low dimensional electron systems continues to attract significant attention in contemporary condensed matter physics. The consequences of this interplay are well established in the fractional quantum Hall regime of two dimensional electron gases. Studies of interaction effects in newer families of materials such as topological insulators are, however, still at a stage of infancy \cite{chiu2016classification,dzero2016topological}. Type-II InAs/GaSb quantum wells provide an attractive platform in this context, with a two-dimensional electron gas (2DEG) and a two-dimensional hole gas (2DHG) spatially separated in the InAs and GaSb layer, respectively, and coupled either by inter-layer tunneling or by the attractive Coulomb interaction \cite{bastard1982,naveh1996,marlow1999}. 

\begin{figure}
\includegraphics{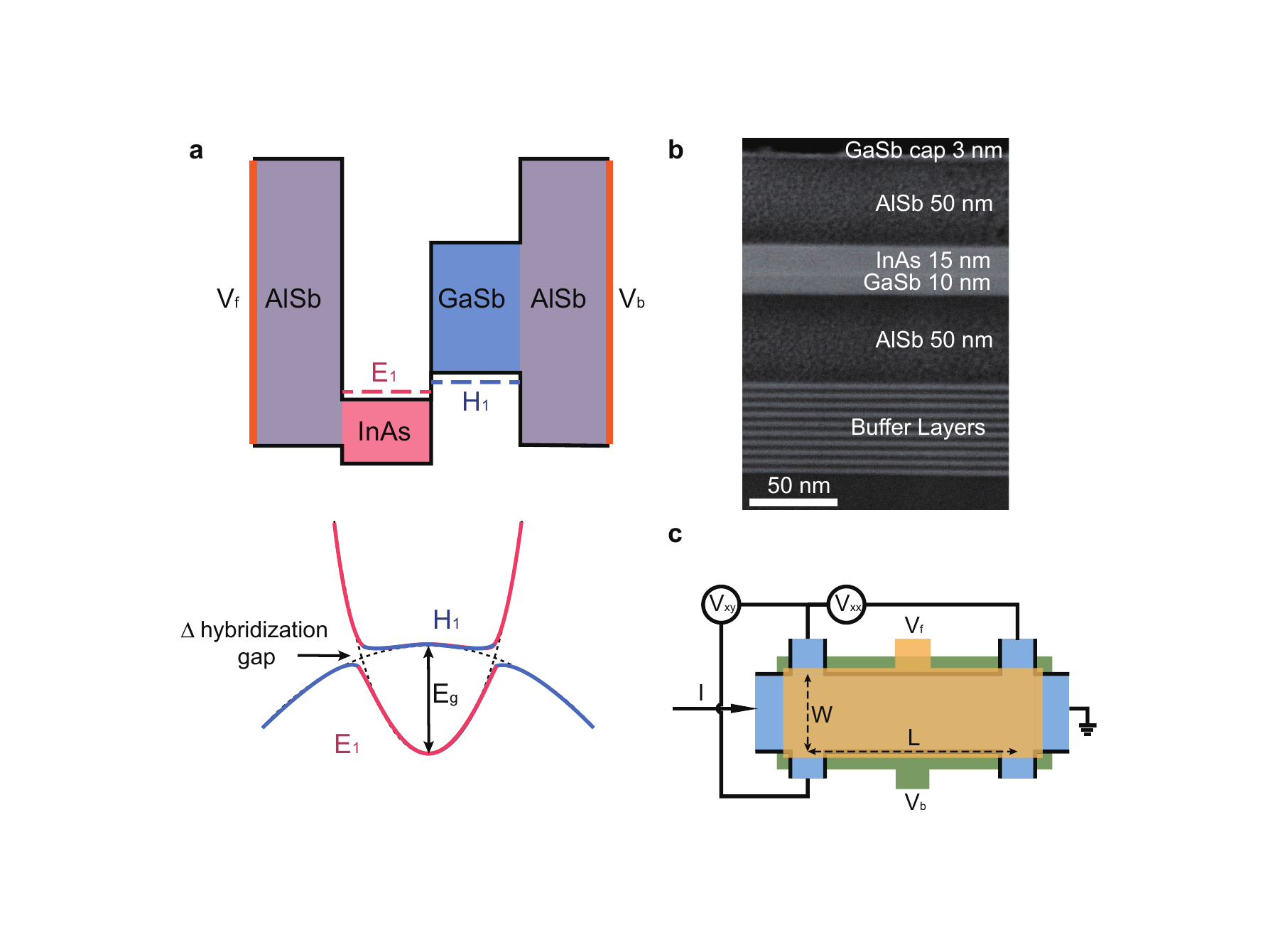}
\caption{\label{fig:eps1} (a) Type-II InAs/GaSb quantum well with inverted band structure, where $\mathrm{E_1}$ is the bottom conduction subband, $\mathrm{H_1}$ is the top valence subband, and $\mathrm{E_g}$ is the negative band gap. (b) Cross-sectional TEM image of a wafer grown by MBE. (c) Schematic of dual-gated Hall measurements.}
\end{figure}

In the non-interacting regime, the combination of the inverted band structure nature of InAs/GaSb quantum wells and the tunneling between the 2DEG and 2DHG leads to a hybridization gap (see Fig.~\ref{fig:eps1}a), resulting in the formation of a 2D topological insulator that might host the quantum spin Hall state \cite{liu2008,knez2011,suzuki2013,du2015}. Extrapolating from earlier studies of interacting 2DEG/2DHG bilayers, one might also anticipate emergent behavior in such InAs/GaSb quantum spin Hall insulators; possibilities include the formation of an exciton insulator \cite{mott1961,jerome1967,kohn1967,halperin1968,bastard1982} or an exciton condensate \cite{blatt1962bose,keldysh1968collective,lozovik1975,snoke2002spontaneous,kasprzak2006bose,PhysRevLett.112.176403,pikulin2016confinement,PhysRevLett.112.146405}
, and the manifestation of the Coulomb drag effect \cite{rojo1999electron,seamons2009coulomb,croxall2008anomalous,nandi2012exciton}. Indeed, recent experimental studies of quantum transport in InAs/GaSb quantum wells \cite{du2015,li2015,du2017evidence} have suggested that the Coulomb interaction might play an important role in determining the physical properties of this 2D topological insulator. For example, experiments suggest the possibility that the repulsive Coulomb interaction between electrons in the helical edge states of an InAs/GaSb 2D quantum spin Hall insulator leads to the formation of a helical Luttinger liquid \cite{li2015}. More recent experiments indicate that the attractive Coulomb interaction between the 2DEG and 2DHG in such a quantum spin Hall insulator results in two regimes with `soft' and `hard' band gaps, with the latter being stabilized by the formation of an exciton insulator, thus making it robust against in-plane magnetic fields\cite{du2017evidence}. The interpretation of these experiments is still evolving and the significance of interaction effects has not been unambiguously established, presumably due to the large single-particle interlayer tunneling. For instance, an alternative scenario argues that the observed excitation gap is a trivial one instead of an inverted one stabilized by an attractive Coulomb interaction \cite{qu2015,nichele2016edge}. Thus, conclusive experiments are still needed for revealing the role of the attractive Coulomb interaction between the 2DEG and 2DHG in InAs/GaSb quantum wells. 

In this Letter, we investigate electrical magneto-transport measurements in $15$ nm InAs/$10$ nm GaSb quantum wells and report the unexpected observation of high frequency Shubnikov de Haas (SdH) quantum oscillations when the chemical potential is tuned close to the charge neutrality regime (corresponding to the hybridization gap). The coexistence of a high sheet resistance (which suggests an insulating behavior) and SdH oscillations with a high oscillation frequency (indicating a high density, high mobility carrier concentration) near the charge neutrality regime suggests that our observations cannot be explained in a non-interacting picture, which is further supported by theoretical modeling. To understand this observation, we develop a heuristic physical picture based on the Coulomb drag between the 2DEG and 2DHG induced by the attractive Coulomb interaction, thus providing strong evidence for the significant role of Coulomb interaction in this system.
 
We grow the wafers with $15$ nm InAs and $10$ nm GaSb on lattice-matched n-GaSb substrates using a standard molecular beam epitaxy technique.\cite{shojaei2018} A representative cross-sectional transmission electron microscopy (TEM) image of a sample is shown in Fig.~\ref{fig:eps1}b; more details about our samples can be found in the supplementary materials. We carry out transport measurements using standard wet-etched Hall bars with $40$ nm $\mathrm{Al_2O_3}$ and $10$ nm$/70$ nm$\ \mathrm{Ti/Au}$ deposited as front gates and conducting epoxy on the bottom of the substrates as back gates (Fig.~\ref{fig:eps1}c). The electron mobility at $400$ mK is about $1.6 \times 10^5$ cm$^2$V$^{-1}$s$^{-1}$ for a density of $1.1 \times 10^{12}$ cm$^{-2}$ without applying bias voltage, comparable with previous studies \cite{du2017evidence,qu2015} Note that the thickness of the InAs/GaSb layers in this study is larger than that in some previous studies, resulting in a larger inversion gap $E_g$. Thus, our samples are in a `deeper' inverted regime, with a higher carrier concentration, compared to the previous samples. 
In addition, the distance between the 2DEG in the InAs layer and the 2DHG in the GaSb layer is larger for a thicker sample, yielding a weaker inter-layer tunneling (hybridization) between the electron and hole gases. 

\begin{figure}
\centerline{\includegraphics{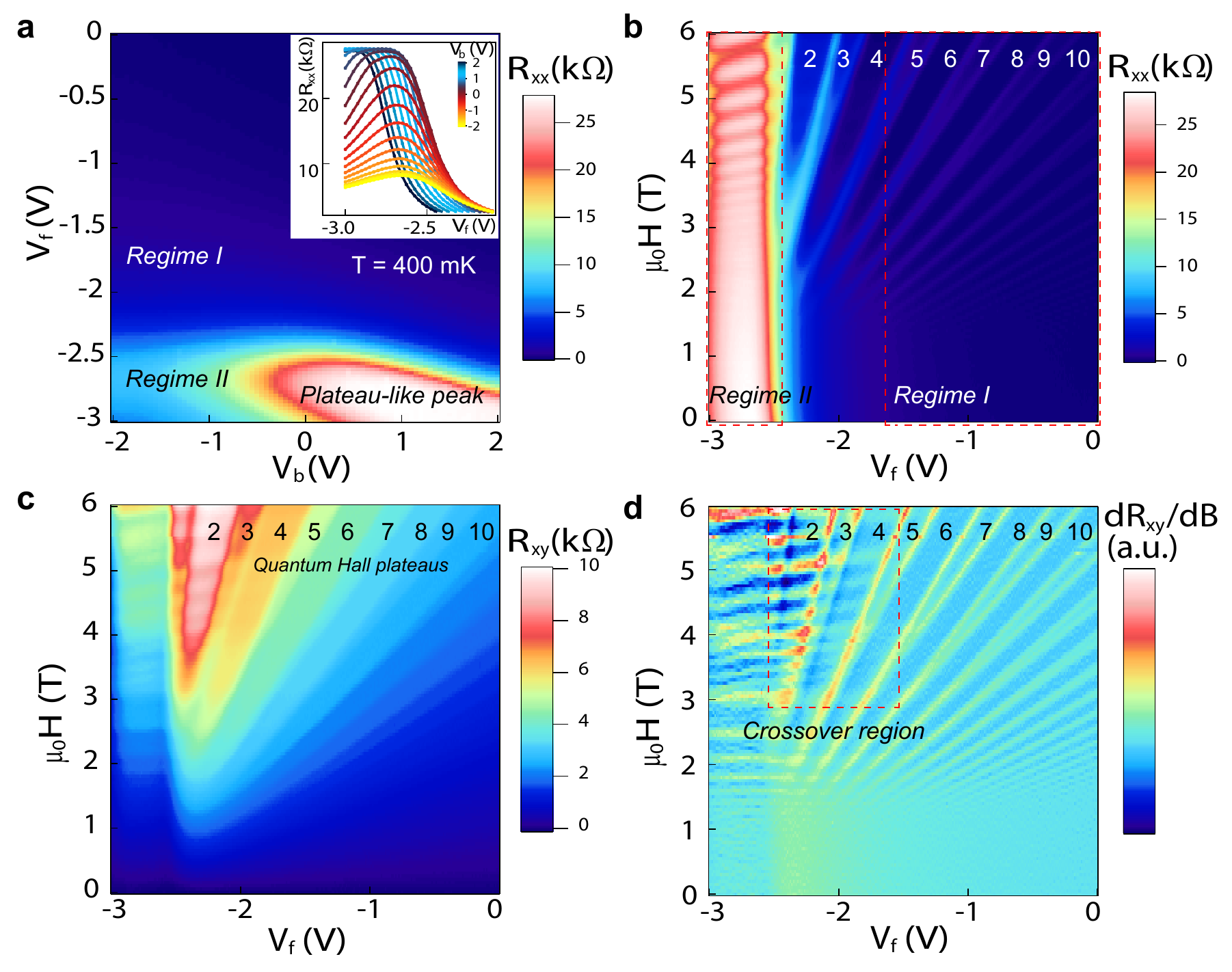}}
\caption{\label{fig:eps2} Measurement of transport in a macroscopic Hall bar. (a) Longitudinal resistance $R_{xx}$ as a function of front and back gate voltages; the inset shows a few vertical line cuts through the color map, plotting the dependence of $R_{xx}$ on $V_f$ at different values of $V_b$ and revealing the resistance plateaus. (b) Dependence of $R_{xx}$ on $V_f$ and magnetic field $B = \mu_0 H$; the red dashed lines demarcate 2 distinct regimes. (c) Hall resistance $R_{xy}$ as a function of $V_f$ and magnetic field $B = \mu_0 H$. (d) Dependence of $dR_{xy}/dB$ on $V_f$ and magnetic field. Two distinct regions are again apparent, one with conducting single-electron-type transport, the other with dense LLs appeared at around $\nu =4$. Red dashed lines indicate the crossover region.}
\end{figure}

Figure~\ref{fig:eps2}a shows the longitudinal resistance $R_{xx}$ of a macroscopic Hall bar ($200\ \mu$m $\times 50\ \mu$m) as a function of the front and back gate voltages (labeled as $V_f$ and $V_b$, respectively). We observe an interesting resistance plateau for positive $V_b$ instead of a sharp resistance peak as previously reported \cite{knez2011,qu2015}
(inset of Fig.~\ref{fig:eps2}a). When $V_b$ is negative, the resistance plateau shrinks into a peak. Figure~\ref{fig:eps2}b plots the longitudinal resistance $R_{xx}$ as a function of $V_f$ and the magnetic field $B$ with $V_b = 0$. This maps out the Landau fan diagram. The corresponding Hall resistance $R_{xy}$ is shown in Fig.~\ref{fig:eps2}c, revealing quantum Hall plateaus with the standard values of $\frac{h}{ne^2}$,
where the integer number $n$ corresponds to the filling factor of Landau levels and can be resolved from 2 to 10. We note two different regimes, I and II, in the Landau fan diagram (Fig.~\ref{fig:eps2}b); these are separated at the critical front gate voltage $V_f \sim -2.0$~V.  Regime I ($V_f  > -1.5$~V) reveals the standard quantum Hall transport behavior, where each peak in $R_{xx}$ corresponds to a plateau transition in $R_{xy}$, thus indicating that the chemical potential crosses a Landau level. This is also seen by plotting $\frac{dR_{xy}}{dB}$ in Fig.~\ref{fig:eps2}d. 

\begin{figure}
\centerline{\includegraphics{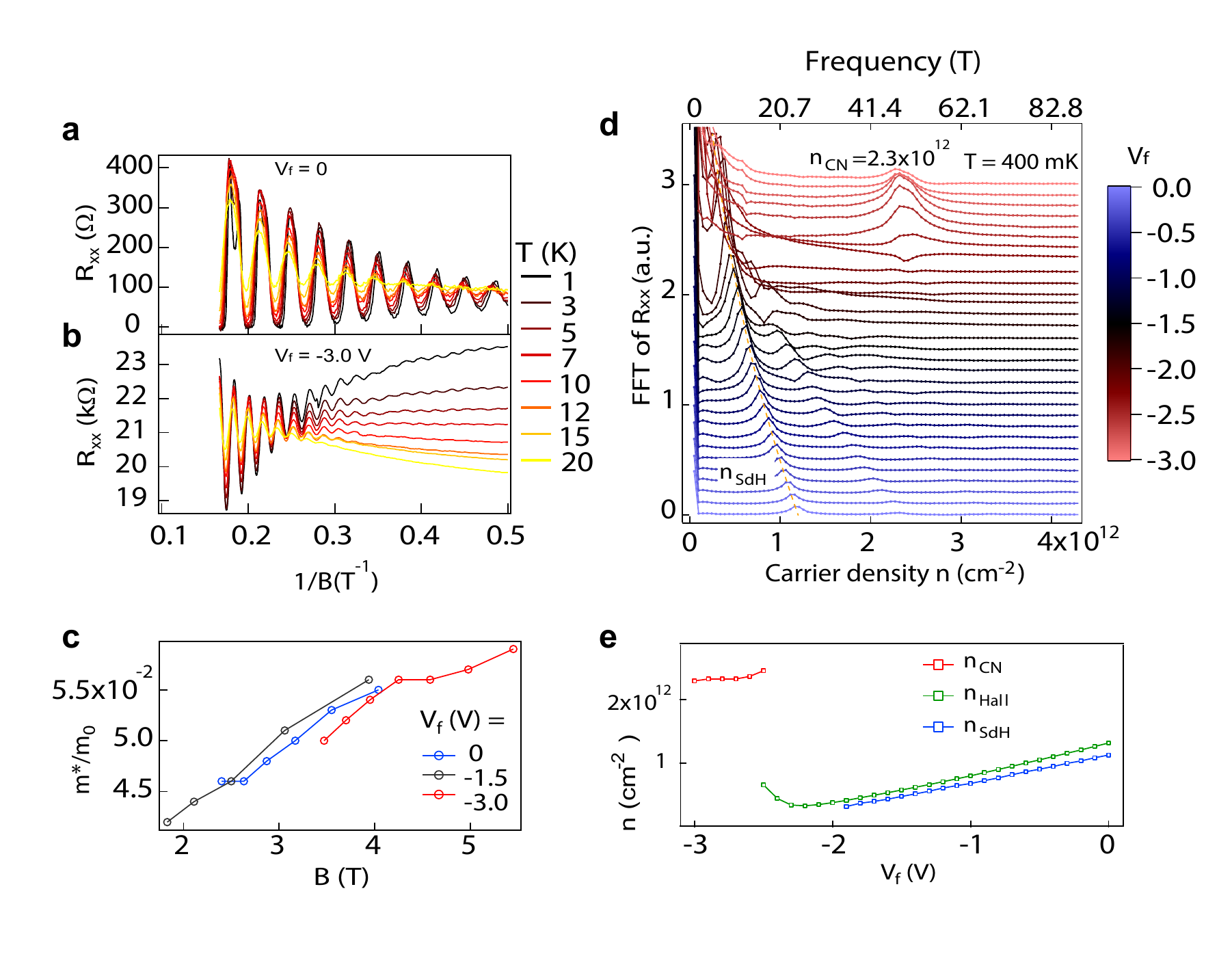}}
\caption{\label{fig:eps3} Analysis of SdH oscillations. (a) $R_{xx}$ as a function of $1/B$ at different temperatures with $V_f=0$~V and (b) $V_f=-3.0$~V. (c) Effective carrier mass extracted from the SdH amplitude. The carriers in the insulating regime have almost identical value as electrons in the InAs conduction band. (d) FFT of $R_{xx}$ vs $1/B$, showing the change of carrier density (and frequency). The carrier density at the charge neutral point is $2.3\times 10^{12}$~cm$^{-2}$. (e) The carrier density calculated from SdH oscillations ($n_{\rm{SdH}}$) and Hall measurements ($n_{\rm{Hall}}$), showing that $n_{\rm{SdH}} < n_{\rm{Hall}}$, especially when the density is high.}
\end{figure}

We note that odd numbers of quantum Hall plateaus are only resolved at a high magnetic field ($> 3.5$~T) when the Zeeman spin splitting is large enough. We can exclude spin splitting due to the Rashba or Dresselhaus contributions \cite{PhysRevB.96.241401} because we do not find any signature of a beating pattern. This is confirmed by a fast Fourier transform (FFT) of $R_{xx}$ versus $1/B$ which yields a single dominant peak in the frequency domain (Figs.~\ref{fig:eps3} (a), (b) and (d)). The absence of any splitting of this peak for different gate voltages indicates that the Fermi surface is almost spin degenerate at zero magnetic field within our resolution, thus further excluding Rashba or Dresselhaus type of spin-orbit effect in our system. 
The additional peak at the higher frequency corresponds to the second harmonic of the oscillation. Accounting for spin degeneracy, the corresponding electron concentration varies from $1.1\times10^{12}$~cm$^{-2}$ to $3.2\times10^{11}$~cm$^{-2}$ when $V_f$ is tuned from $0$ to $-2$~V. This electron concentration is a bit smaller than that extracted from Hall measurements for $-2$~V$ < V_f <0$~V, as shown in Fig.~\ref{fig:eps3}e. The effective mass of electron carriers can be extracted from the temperature dependence of the oscillation amplitude, and ranges from $0.04 m_0$ to $0.055 m_0$, similar to values obtained in a previous study \cite{mu2016}. The magnetic field dependence of effective mass can be explained by the non-parabolic character of the conduction band \cite{palik1961}.

The most striking observation in our experiments is the presence of high frequency quantum oscillations in regime II for the front gate voltage $-3.0$~V$< V_f < -2.5$~V (Fig.~\ref{fig:eps2}b, c and d). 
The rapid increase of resistance at zero magnetic field (see the inset in Fig.~\ref{fig:eps2}a) suggests that the system is tuned into the hybridization gap regime. This is further shown by the temperature dependence of the longitudinal resistance, which reveals an insulating behavior for temperature above $2$ K and saturates around $28\ \mathrm{k}\Omega$ when the temperature is below $1.5$ K, which indicates potential existence of in-gap states (See supplementary materials for details). 
This oscillation is also periodic in $1/B$, as shown in Fig.~\ref{fig:eps3}b, thus suggesting that it originates in the SdH effect. 
The Fourier transform of this oscillation yields a carrier concentration $2.3\times10^{12}$cm$^{-2}$, 
which is almost independent of the front gate $V_f$ (Fig.~\ref{fig:eps3}d). 
Compared with the electron concentration around $3.2\times10^{11}$cm$^{-2}$ at the gate voltage $V_f = -1.9$~V, 
this carrier concentration is about 7 times larger and thus we find an abrupt jump in carrier concentration as a function of $V_f$, as shown in Fig.~\ref{fig:eps3}e. 
In the plot of $R_{xx}$ as a function of top gate voltage and magnetic field (Fig.~\ref{fig:eps2}b), regimes I and II seem well separated. However, we note that if we plot $\frac{dR_{xy}}{dB}$, there is actually a crossover regime for
the gate voltage $-2.5$~V$ < V_f < -1.5$~V, where both high and low frequency quantum oscillations 
coexist. The effective mass for the carriers in  regime II can also be extracted from 
the temperature dependence, as shown in Fig.~\ref{fig:eps3}c, and is slightly smaller than that in regime I. 

At the first sight, the transport behavior in the regime II is puzzling because the large resistance (insulating behavior) suggests the absence of free carriers and is in contradiction with the high frequency of the quantum oscillations: the latter indicates a high free carrier concentration. We note that recent theoretical calculations for the inverted band structure regime have predicted the existence of quantum oscillations even when the chemical potential lies in the hybridization band gap \cite{knolle2015,zhang2016}. Evidence for such quantum oscillations is suggested in the de Haas-van Alphen effect in topological Kondo insulators \cite{tan2015}. Recently, transport evidence of SdH quantum oscillations has also been reported in related materials.\cite{xiang2018} However, the influence of the attractive Coulomb interaction on quantum magneto-transport experiments is still unclear \cite{pikulin2014,knolle2017anomalous,knolle2017,erten2016,chowdhury2018}. According to these theoretical predictions, quantum oscillations in the inverted band gap regime are controlled by the area $A_{\rm{edge}}$ enclosed by the band edge, $F=\frac{\hbar}{2\pi e}A_{\rm{edge}}$ \cite{zhang2016}(Fig.~\ref{fig:eps1}a). Therefore, we should expect a continuous change of the oscillation frequency when the chemical potential is tuned across the hybridization gap since the area should vary continuously with varying chemical potential. Indeed, our calculations of the oscillation of density of states in a non-interacting four-band model with inverted band structure under magnetic fields also predict a continuous change of the quantum oscillation frequency (see Supplementary Material). Thus, the current theoretical model fails to explain the abrupt jump of oscillation frequency observed in our experiments when tuning the gate voltage from regime I to regime II. Further, we note that the extremely high frequency of the quantum oscillations in regime II is inconsistent with the model. 
 
We attribute this failure of the theoretical model to the neglect of interactions. We emphasize that the attractive Coulomb interaction can lead to the binding between the 2DEG and 2DHG in the inverted band structure, thus leading to a strongly interacting electron-hole gas in our system. The coexistence of two different frequency oscillations in the gate voltage regime $-2.5 \lesssim V_g \lesssim -1.5$ V in Fig.~\ref{fig:eps2}d and Hall data in Supplementary Material further implies 
 two parallel transport channels existing in our systems. 
 
\begin{figure*}
\centerline{\includegraphics{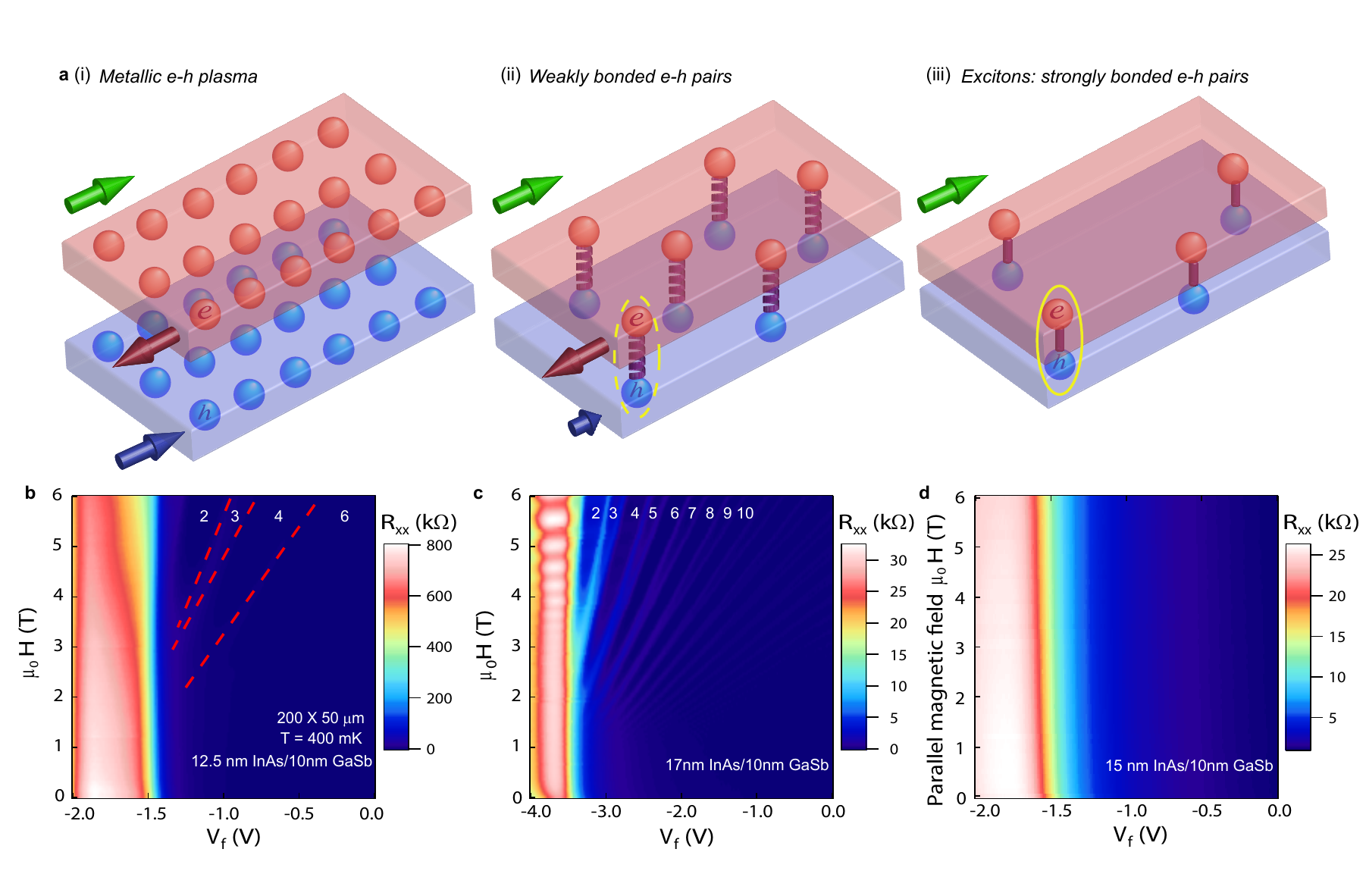}}
\caption{\label{fig:eps4}Schematic illustration of (i) an electron-hole plasma with no interactions, (ii) weakly coupled e-h pairs, and (iii) strongly coupled e-h pairs. Under an external electric field, these behave like (i) a metal, (ii)a `quasi-exciton' insulator and (iii) an exciton insulator, respectively. $V_f$ and magnetic field dependence of $R_{xx}$ of (b) a macroscopic 12.5 nm InAs/10 nm GaSb sample and (c) a macroscopic 17 nm InAs/10 nm GaSb sample. (d) Variation of $R_{xx}$ in a macroscopic 15 nm InAs/10 nm GaSb sample as a function of $V_f$ and in-plane magnetic field.}
\end{figure*}

We use a heuristic model with two transport channels, one channel involving a free 2DEG with concentration corresponding to the peak labeled by $n_{\rm{SdH}}$ in Fig.~\ref{fig:eps3}d, e, and the other described by bound electron-hole pairs (excitonic quasi-particles) with the concentration corresponding to the peak $n_{\rm{CN}}$ in Fig.~\ref{fig:eps3}d and e. 
 In regime I, the transport is dominated by free electron carriers, which give rise to the standard quantum Hall effect.
 In regime II, free electrons are depleted and thus only excitons remain. One may ask how 
 excitons can participate in the transport process since they are charge neutral. 
 We make the assumption that the binding energy of excitons in our system is quite weak, 
 and thus the applied electric field can ionize the excitons into electron-hole pairs and 
 lead to charge transport. An alternative physical picture is to regard this exciton ionization process 
 as a Coulomb drag of coupled electron-hole gases, as depicted in Fig.~\ref{fig:eps4}a.
 Considering an electron-hole bilayer without Coulomb interaction, electrons move along the electric field
 while holes move in the opposite direction. Since electrons and holes carry opposite charges, a huge current will
 be driven, giving rise to a metallic behavior (Fig.~\ref{fig:eps4}a(i)). 
 In the opposite limit, if electrons and holes are strongly bound
 and move together, no electric current can be induced and thus we obtain an exciton insulator (Fig.~\ref{fig:eps4}a(iii)).
 Our system stays in the intermediate regime, where the Coulomb interaction couples electron and hole gases, 
 but not that strongly. As a result, holes move with electrons but with a different velocity, thus greatly reducing 
 electric current (insulating behavior). However, net charge transport still remains; thus, the quantum oscillations in the insulating regime become possible. This corresponds to the Coulomb drag picture. The high frequency of the oscillation suggests a high density coupled electron-hole gas in our system. This physical picture is also consistent with the following experimental observations: first, we observe a reduced effective mass in regime II, consistent with a smaller effective exciton mass compared to that of electrons; second, the discrepancy between $n_{\rm{Hall}}$ and $n_{\rm{SdH}}$ in Fig.~\ref{fig:eps3}e is consistent with the influence of a high exciton concentration $n_{CN}$ although it should be mainly screened by free electrons. 

To further support our viewpoint, we present additional control experiments with different InAs layer thicknesses. 
In a $12.5$ nm InAs/$10$ nm GaSb sample, the quantum oscillation disappears in the gap regime
and the corresponding peak resistance is over $800\ \mathrm{k}\Omega$ at zero magnetic field,
as shown by the LL mapping in Fig.~\ref{fig:eps4}b. In contrast, quantum oscillations exist 
in a $17$ nm InAs/$10$ nm GaSb sample, as shown in Fig.~\ref{fig:eps4}c. 
For the thinner samples, both inter-layer tunneling and attractive Coulomb interaction between 
electron and hole gases are enhanced, thus leading to the suppression of quantum oscillations. 

Finally, we study the dependence of $\mathrm{R_{xx}}$ on an in-plane magnetic field (perpendicular to the Hall channel). Figure ~\ref{fig:eps4}d shows the corresponding data from a different sample (but from the same wafer) with $15$ nm InAs/$10$ nm GaSb and the same device size. In contrast with past experiments showing an obvious shift and reduction of the resistance peak at high fields\cite{qu2015,shojaei2018,karalic2016}, 
our sample has a weak dependence ($\sim5\%$ decrease) on the in-plane magnetic field up to $6$ T.
This provides further support that the insulating behavior is induced by an attractive Coulomb interaction, instead of the hybridization effect\cite{du2017evidence}. 

To conclude, we have studied quantum magneto-transport in InAs/GaSb quantum wells in the deeply inverted regime. 
The observation of high frequency quantum oscillations in the charge neutrality regime
provides strong evidence of attractive Coulomb interaction between electron and hole gases,
suggesting an interesting interplay between the Coulomb interaction and interlayer tunneling in this topological system. 
Our surprising observation calls for more quantitative theoretical studies 
of this new phase of strongly interacting electron-hole gases produced by a combination 
 of an inverted band structure and Coulomb interaction. 
\begin{acknowledgments}
D.X. and N.S. acknowledge support from Office of Naval Research (Grant No. N00014-15-1-2370) and from ARO MURI (W911NF-12-1-0461). C.X.L. acknowledges the support from Office of Naval Research (Grant No. N00014-15-1-2675).
\end{acknowledgments}

%

\end{document}